\documentclass[reprint, amsmath, amssymb, aps, prl]{revtex4-1}
\usepackage[utf8]{inputenc}
\usepackage{braket}
\usepackage{graphicx}
\usepackage{xcolor}
\usepackage{subcaption}

\begin{document}

\title{Simulation of 1D topological phases in driven quantum dot arrays}

\author{Beatriz P\'erez-Gonz\'alez}
\email{bperez03@ucm.es} 
\author{Miguel Bello} 
\author{Gloria Platero}
\author{\'Alvaro G\'omez-Le\'on} 
\affiliation{Instituto de Ciencia de Materiales de Madrid (ICMM-CSIC)}
\date{\today}

\begin{abstract}
We propose a driving protocol which allows to use quantum dot arrays as quantum simulators for 1D topological phases. We show that by driving the system out of equilibrium, one can imprint bond-order in the lattice (producing structures such as dimers, trimers, etc) and selectively modify the hopping amplitudes at will. Our driving protocol also allows for the simultaneous suppression of all the undesired hopping processes and the enhancement of the necessary ones, enforcing certain key symmetries which provide topological protection. In addition, we have discussed its implementation in a 12-QD array with two interacting electrons and found correlation effects in their dynamics, when configurations with different number of edge states are considered.
\end{abstract}


\maketitle

\textit{Introduction}: Topological matter, and in particular topological insulators (TIs) \cite{KaneMele, ColloqiumTI, TIandSC} are materials of interest due to the presence of topologically protected surface states, robust to local perturbations. Consequently, a great effort is being made to simulate the behaviour of TIs by tailoring other quantum systems, whose properties can be more easily controlled. Within this context, time-dependent modulations have proven to be useful tools to modify  the topology \cite{Miguel1, Miguel2, pwavesupercond, beyondpwavesupercond, topcurrentblockade, charlesgloria, charles, quantumrouter, FloquetKlinovaja, FMajorana, AguadoGloria, floquetmajoranaDiptiman}. Particularly, they have been used to simulate the so-called "Floquet topological insulators" (FTIs) \cite{ftiCyassol, FTILindnerRefael, photovoltaicgraphene} upon different systems \cite{photovoltaicgraphene, floquetgraphene, tunableTIgraphene, floquetbilayergraphene, floquetgrapheneribbon, floquetdelplace, MariaGloria, drivenhoneycomb, AlvaroHoneycomb, alvarodrivenssh, photoinduced2D, floquet3D, drivenSSHlowfreq, modulatedFTI, floquetcoldatoms, floquetsymmetrycoldatoms, shakingopticallattices,photonicFTI, obvAFTIsound2, obvAFTIsound, heatingRef, heatingFloquet}.\\
Quantum dots (QDs) have revealed themselves as highly tunable quantum systems \cite{Petta2180, Petta2,controlEqd, controlEqd2}, in which both on-site energies \cite{onsiteQD} and couplings \cite{tunnelingQD1, tunnelingQD2} can be independently addressed. This makes them an interesting platform for quantum simulation \cite{simquantumdot, QDchemicalreactions, QDsimulator}. Recent experimental evidence on scalable quantum dot devices \cite{petta9qubits, vandersypen8} demonstrates reproducible and controllable long QD arrays, which opens up new possibilities of simulating 1D TIs.\\
In this work we show that a quantum simulator for 1D topological phases can be obtained by periodically driving an array of QDs with long-range hopping. We propose a driving protocol which allows us to imprint bond-order in the lattice \cite{chinosDimersTrimers}, while also offers tunability for the long-range hoppings. This control can lead to configurations that would be unreachable otherwise, while preserving the fundamental symmetries which guarantee topological features. Thus, the driving protocol triggers non-equilibrium topological behavior in a trivial setup, opening the door to the simulation of different topological phases. We also study the exact time-evolution for the case of two interacting electrons, and show that the dynamics of different edge states can become highly correlated. This allows to discriminate between different topological phases and also opens up new possibilities for quantum state transfer protocols. \\
Our proposal can also be implemented in other set-ups as cold atoms or trapped ions \cite{Gauthier16,Stuart2018,opticalAddressing,tunnelingOL,measurementZak,experimentoSSHcita, topOLexp,topOLexp2}.

\textit{Model}: We consider a Hamiltonian describing a periodically driven chain of QDs:

\begin{equation}
  \begin{split}
    H(t) & = \sum_{|i-j|\leq R}J_{ij}c^{\dagger}_i c_j 
    + \sum_{i} A_i f(t) c^{\dagger}_i c_i \\
    & \equiv H_{\mathrm{array}} + H_{\mathrm{\mathrm{driv}}}(t)\,,
  \end{split}
  \label{eq:Hamiltonian}
\end{equation}

where $c^{\dagger}_i$ ($c_i$) is the creation (destruction) operator for a spin-less
fermion at the $i^{\mathrm{th}}$ site of the array. The first term represents
the static Hamiltonian for a QD array of $N$ sites, with $J_{ij}$
being the real hopping amplitude connecting the $i^{\mathrm{th}}$ and
$j^{\mathrm{th}}$ dots. Note that long-range hoppings are allowed to take place,
up to range $R$. We will assume that hopping amplitudes in the undriven system decay monotonically as a
function of the distance between sites, $J_{ij}=J(|i-j|)$. The
second term in Eq.~\eqref{eq:Hamiltonian}, $H_{\mathrm{\mathrm{driv}}}(t)$,
corresponds to a time-periodic modulation of the on-site potentials $A_i$ with 
$f(t)=f(T+t)$, and frequency $\omega=2 \pi/T$. 

Regarding the simulation of 1D topological phases in QD arrays, the purpose of the time-periodic modulation is threefold: First, the driving must generate bond order, which is a crucial ingredient in toy models such as the SSH \cite{SSHmodel}. Second, certain neighbor hoppings can be simultaneously suppressed through the so-called coherent destruction of tunneling \cite{miprimerarticulo,coherentdestruction1}. This difficult requirement turns out to be feasible when our driving protocol is included. Finally, other hoppings can be enhanced in order to generate the necessary symmetries for the topological protection, and to be able to explore topological sectors with larger topological invariant.

All these objectives can be achieved through a spatially modulated square ac field \cite{squarewaveCref,squarewave2,squarewave3}

\begin{equation}
 f(t) = \left\{
 \begin{array}{lcr}
 -1 & \mathrm{if} & 0\leq t <T/2\\
 1 & \mathrm{if} & T/2 \leq t<T
 \end{array} \right. \,,
\end{equation}

and in particular, the simulation of an effective dimer lattice with long-range hopping can be realized by choosing the $A_i$ in a stair-like fashion,

\begin{equation}
A_{2n}=n(\alpha+\beta),\hspace{5pt}A_{2n-1}=n(\alpha+\beta )-\alpha.
\end{equation}

with $n = 1,2,3,...$, which translates into an alternating difference between two consecutive sites, namely $A_{2n}-A_{2n-1} = \alpha $ and $A_{2n+1}-A_{2n} = \beta$.
Given the time-periodicity of the Hamiltonian $H(t)=H(t+T)$ \cite{floquettheory,
FloquetTheoryGoldman}, we can take advantage of Floquet theory to solve the
time-dependent Schrödinger equation. The solutions take the form
$\ket{\psi_n(t)}=e^{-i\epsilon_n t}\ket{u_n(t)}$, where the so-called Floquet
modes $\ket{u_n(t)}=\ket{u_n(t+T)}$ have the same periodicity as the
Hamiltonian, and $\epsilon_n$ are the so-called quasienergies, which play an analogous role to the energies in the static Hamiltonians. In the high-frequency regime ($\omega\gg J_0$), the dynamics is essentially dictated by the stroboscopic evolution of an effective time-independent Hamiltonian $H_{\mathrm{eff}}$, which can be derived with a Magnus expansion. This leads to an effective Hamiltonian identical to $H_{\mathrm{array}}$, but with renormalized hopping amplitudes \cite{supp}:

\begin{equation}
J_{ij}\rightarrow \tilde{J}_{ij}=J_{ij}\frac{i\omega }{\pi(A_i-A_j)}\bigg( e^{-i\pi (A_i-A_j)/\omega}-1 \bigg)\,.
\label{renorm}
\end{equation}

From Equation \eqref{renorm} we can see that even-neighbor hoppings $J_{i,i\pm 2m}$ with $m=1,2,3,...$ ($\pm$ for hoppings to the right and left, respectively) renormalize through $A_{i}-A_{i\pm 2m}=\mp m(\alpha + \beta )$. This is important, because topological phases with chiral symmetry can be spoiled by the presence of hoppings connecting sites within the same sublattice. The quenching of all $\tilde{J}_{i,i\pm 2m}$ can be achieved by choosing $\alpha+\beta = 2\omega q$, with $q=0,1,2,...$. Hence, chiral symmetry is recovered, independently of the maximum range of the hoppings included. 

On the other hand, the renormalization of odd-neighbor hoppings leads to bond ordering, due to the alternating structure of the driving protocol. Together with the presence of chiral symmetry, this ensures the existence of distinct topological phases. We identify the renormalized $J_{2i,2i-r}$ as $\tilde{J}_{-r}'$ and $J_{2i+r,2i}$ as $\tilde{J}_{r}$ ($r \in [1,3,5,\ldots,R]$), obtaining \cite{supp}

\begin{equation}
\begin{split}
\tilde{J}_{-r}'&=\frac{iJ_{2i,2i-r}}{\pi \big[\frac{\alpha}{\omega} + (r-1)q \big]}\bigg[e^{-i\pi\big( \frac{\alpha}{\omega} + (r-1)q\big)} - 1 \bigg]\,,\\
\tilde{J}_r&=\frac{iJ_{2i+r,2i}}{\pi\big[ (r+1)q-\frac{\alpha}{\omega} \big]}\bigg[e^{-i\pi\big((r+1)q-\frac{\alpha}{\omega} \big)} - 1 \bigg]\,.
\end{split}
\label{renorm1}
\end{equation}

Notice that now long-range odd-hoppings can be tuned, while keeping even-hoppings suppressed. This can make long-range hoppings dominate over short-range ones, and then allows to explore different topological phases by just tuning the driving amplitudes. The sign of $r$ in the subscript is relevant since hopping amplitudes are now complex functions, and hence $\tilde{J}^{(\prime)}_{\pm r}=(\tilde{J}^{(\prime)}_{\mp r})^*$.

Interestingly, our protocol can be generalized to reproduce different kinds of bond-ordering and to enforce other symmetries as well by choosing the driving on-site amplitudes accordingly. A trimer chain \cite{trimerchain} is an particular example of a 1D system hosting non-trivial topological phases that can be realized in our set up. In this case, chiral symmetry is intrinsically absent, but the presence of another crystalline symmetry, space-inversion symmetry, can provide for topological protection \cite{invsymtopology}. A trimer chain can be realized in a QD driven monomer chain just by considering $A_{2n} - A_{2n-1} = A_{2n+1} - A_{2n} = \alpha$ and $A_{2n + 2} - A_{2n + 1} = \beta$. \\

\begin{figure}
	\includegraphics[width=\linewidth]{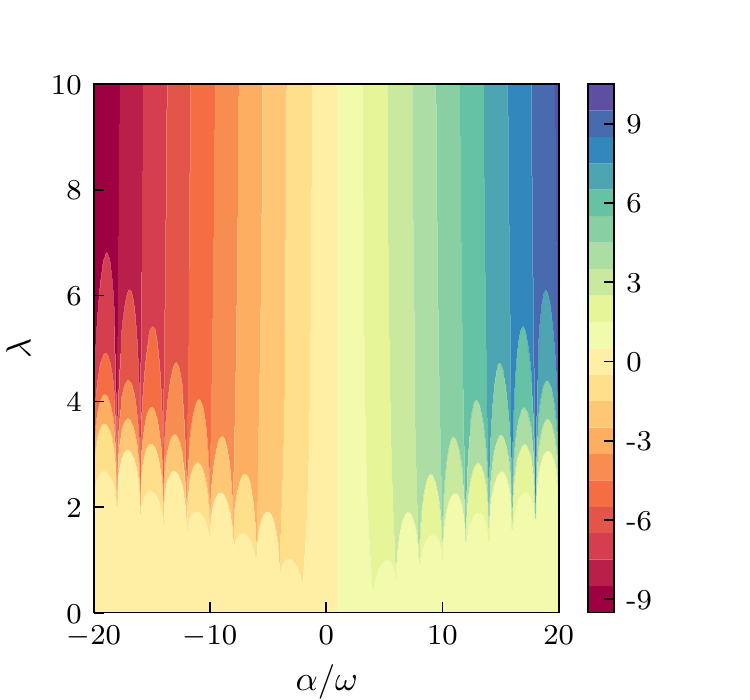}
	\caption{\label{fig:phasemap} Topological invariant $\mathcal{W}$ as a function of the driving amplitude $\alpha/\omega$ and decay length of hopping amplitudes $\lambda$, for $q=1$. The maximum range of the hoppings included is fixed by $\lambda$, such that the longest bare hopping amplitude is a factor $10^{-8}$ smaller than the shortest one.}
\end{figure}

\textit{Topological phase diagram for driven QD arrays}. In QD arrays, the bare hopping amplitudes typically decay exponentially with distance, with a decay length $\lambda$: $J_{ij}=J e^{-d_{ij}/\lambda}$, where $d_{ij}$ is the distance between the $i^{\mathrm{th}}$ and $j^{\mathrm{th}}$ dots and $J$ is of the order of tens of $\mu \mathrm{eV}$, which are the typical energy scales in these setups. The distance between two consecutive QDs is set to $a=1/2$ so that the unit cell in the effective dimerized chain is $1$. By varying the value of $\alpha$ and $\lambda$ in the driven system, topological phases with different topological invariant can be realized (Fig.\ref{fig:phasemap}). The topological invariant $\mathcal{W}$ is calculated as the winding number of the Bloch vector $\vec{d}(k)=(\mathrm{Re}[d(k)],-\mathrm{Im}[d(k)])$ around the origin \cite{delplace}, assuming a system with periodic boundary conditions, with $d(k)$ defined as

\begin{equation}
d(k) = \sum_{r=1}^R \bigg[ \tilde{J}^{\prime}_{-r} e^{i k \frac{r-1}{2}}+ \big(\tilde{J}_{r} \big)^{*} e^{-i k \frac{r+1}{2}} \bigg],
\end{equation}

For small values of $\lambda$, only $\mathcal{W}=1$ and $\mathcal{W}=0$ phases are allowed for all values of $\alpha/\omega$, since first-neighbor hoppings are dominant (this corresponds to the SSH model). Then, when $\lambda$ is increased, other phases with larger $\mathcal{W}$ are possible, as a function of the ratio $\alpha/\omega$ (we have also calculated the size of the gap in the Supplemental Material \cite{supp}, as one is typically interested in gap sizes smaller than the temperature of the setup).

Typically, other driving protocols have been considered in the literature, such as sinusoidal driving fields $f(t) =\sin(\omega t +  \phi)$ \cite{alvarodrivenssh, quantumrouter, charles, topcurrentblockade}, or standing waves $f(t)= \cos(\omega t)\sum_{j=1}^{N}\cos(k x_j + \phi)$ \cite{diegoporrasDriving}. However, none of them would be suitable for engineering arbitrary chiral topological phases. In both previous cases, renormalization of the hopping amplitudes occurs through a zero-order Bessel function, whose roots are not periodically spaced. Hence, it would not be possible to suppress all even hoppings at once, and chiral symmetry would not be present. On the other hand, a homogeneous square driving field with $A_i = A$ could restore chiral symmetry in a dimerized chain but cannot generate topological phases beyond those with $\mathcal{W}=0,1$ if hoppings decay exponentially with distance.

\begin{figure}
	\includegraphics[width=1.1\linewidth]{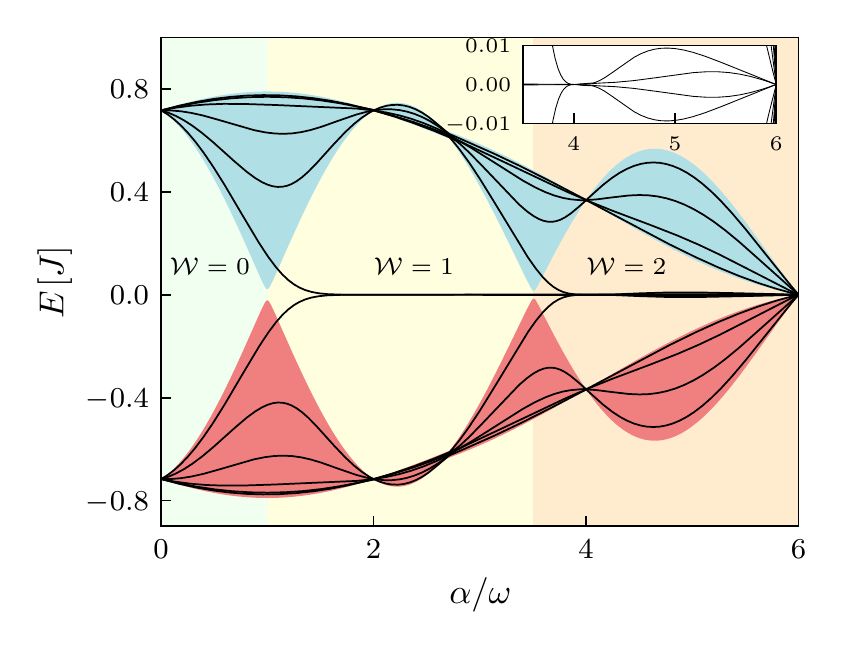}
	\caption{Quasienergy levels of a driven 12-quantum dot array and band structure in the thermodynamical limit (red and blue filling for the valence and conduction band, respectively), as a function of $\alpha/\omega$, including first- and third- neighbor hoppings, in the high-frequency regime. Second-neighbor hoppings have been suppressed through the driving protocol. The parameters are $\lambda/\omega = 1.5$, $q=1$. Inset: each pair of edge states in the $\mathcal{W}=2$ phase has a different energy splitting, which can be also varied by tuning $\alpha/\omega$.}
	\label{fig:quasienergies}
\end{figure}

The experimental evidence provided in \cite{petta9qubits} demonstrates a reproducible and controllable 12-quantum-dot device. Motivated by this experimental setup, we propose the implementation of our driving protocol in an array of 12 quantum dots. 
In Fig.\ref{fig:quasienergies} we show the quasienergies, as given by the effective Hamiltonian, of a driven 12-quantum-dot array, as a function of $\alpha/\omega$, with first- and third-neighbor hoppings (second-neighbor hoppings were initially present, but are effectively suppressed by the driving protocol), fixing $\lambda=1.5$. The spectrum shows two topological phases with $\mathcal{W}=1$ and $\mathcal{W}=2$.

\textit{Dynamics of two interacting particles}. The number of edge states hosted by a finite system and their localization properties determine the motion of charges along the chain. Then, for an electron initially occupying the ending site, one would see oscillations between the two edges of the chain, with a frequency defined by the energy splitting: $\omega_{\mathrm{osc}}\propto E_{+}-E_{-}$, being $E_{\pm}$ the energy of each edge state in the pair. Hence, one can discriminate between topological phases with different number of edge states by studying the electron dynamics.

\begin{figure*}
	\centering
	\includegraphics[width=\linewidth]{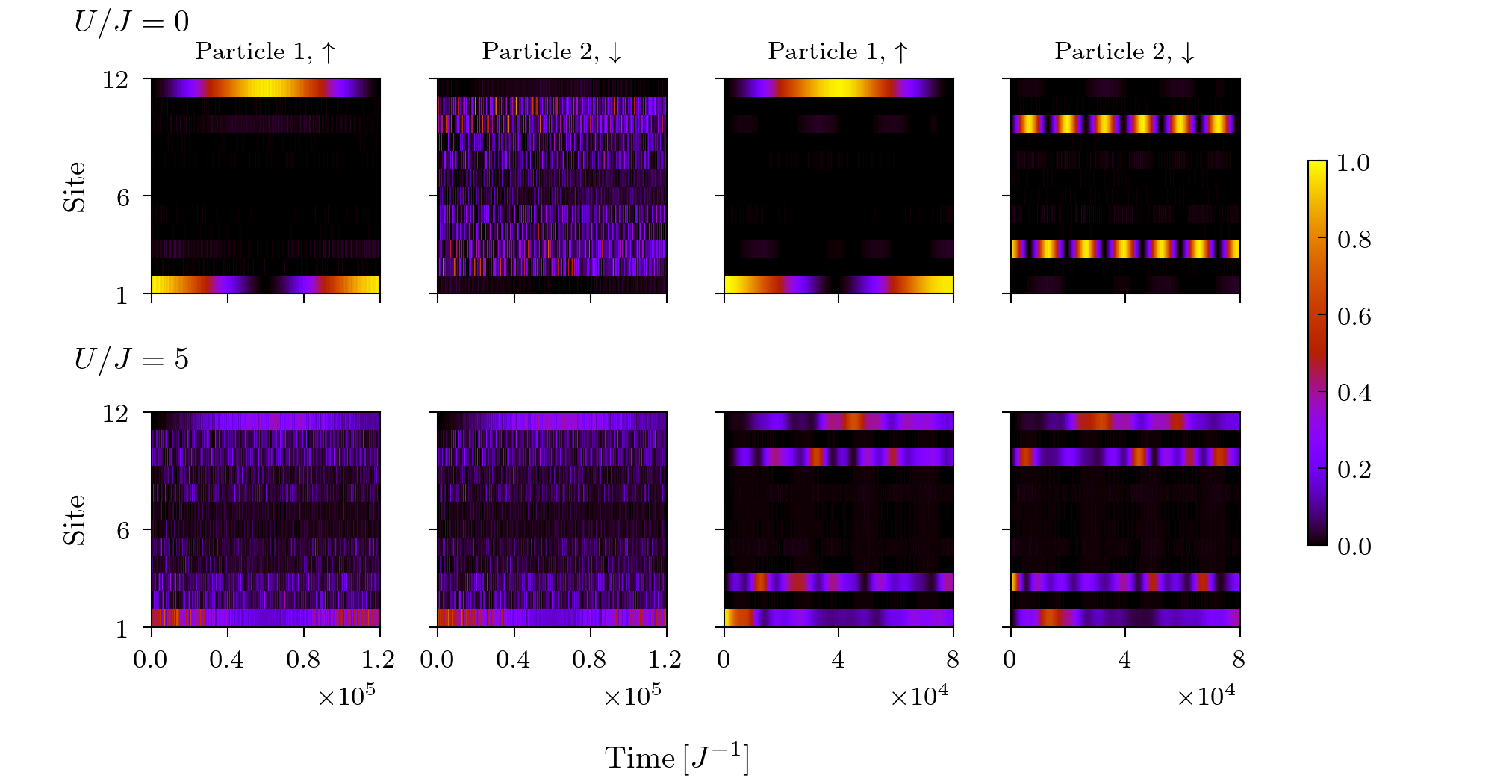}
	\caption{Occupation of each site of a driven 12-QD array as a function of time when two electrons with opposite spin are loaded into the system, for different values of $U$. The array has first- to third-neighbor couplings, whose values initially decay exponentially with distance, choosing $\lambda = 1.5$. Considering the high-frequency regime with $\omega=100$ and $q=1$, the values for $\alpha$ have been chosen such that the desired hopping renormalization is realized. The four left plots correspond to $\alpha=230$, resulting in a topological phase with $\mathcal{W}=1$ (two edge states). The four right plots correspond to $\alpha=410$, yielding a configuration with $\mathcal{W}=2$ (four edge states) \label{fig:dynamics}}
\end{figure*}

These ideas are illustrated in Fig.\ref{fig:dynamics}, where we consider two electrons with opposite spin loaded in a driven array of 12 QDs in such a way that the spin-up (spin-down) particle, which we will denote as $\uparrow$ ($\downarrow$), initially occupies the first (third) site. We have also included a local interaction term, being the total Hamiltonian:
\begin{equation}
\begin{split}
H_{\uparrow \downarrow }(t) &= \sum_{\sigma}\sum_{|i-j|\leq R}J_{ij}c^\dagger_{i,\sigma}c_{j,\sigma} + \sum_{i,\sigma}A_i f(t)n_{i,\sigma}\\
& + U\sum_{i}n_{i,\uparrow}n_{i,\downarrow},
\end{split}
\end{equation}
where $\sigma=\uparrow,\downarrow$. We do not include any spin-flip terms, since experimental evidence on silicon QDs confirms that the spin relaxation time within these QD structures is very long compared with the other energy scales of the system \cite{spinflipref}. The $A_i$ are chosen as indicated before, and $\alpha$ is fixed such that the system hosts either two or four edge states. 

Then, the dynamics is exactly calculated from the time evolution operator $U(t,0)=e^{-i\int^t_0 H_{\uparrow \downarrow}dt}$. Since $H_{\uparrow \downarrow}$ is time-independent in each half period, the time-evolution operator $U(T,0)$ can be factorized into two independent time-evolution operators, $U(T,0)=U(T)=U_{+}(\frac{T}{2})U_{-}(\frac{T}{2})$, where the subscript $\pm$ corresponds to the sign of $f(t)$ in each of them. We choose for our simulations $\omega\gg J$ in order to accurately match the analytic expression in Eq.\eqref{renorm1}, however we have checked that values $\omega \gtrsim J$ still produce the expected behavior.

First, $\alpha$ is chosen such that the system hosts one pair of edge states ($\mathcal{W}=1$ for the left half of Fig.\ref{fig:dynamics}), which have the largest weight at the ending sites of the chain. When interaction is turned off, particle $\uparrow$ oscillates between the ends of the chain, while particle $\downarrow$ spreads along the chain: at the third site, other states from the bulk have a non-negligible contribution and the edge states do not dominate the dynamics. When $\alpha$ is fixed so that the system has four edge states ($\mathcal{W}=2$ for the right half of Fig.\ref{fig:dynamics}), one of the pairs is maximally localized at the first and last sites, while the other has the largest weight at the third and second-to-last sites. Hence, each particle is coupled to a different pair of edge states and it displays oscillations between different sites. The frequency of oscillation is also different, since each pair has a different energy splitting (see inset in Fig.\ref{fig:quasienergies}). The second pair has a bigger splitting and thus oscillations for particle $\downarrow$ happen faster.

Interestingly, for the case of non-vanishing local interaction one can see that the general effect is to correlate the dynamics of the two electrons. For the case of just one pair of edge states, the interaction correlates the edge mode with the bulk. They exchange spectral weight and oscillate coherently. However the case of two pairs of edge modes is more interesting, as the interaction correlates their dynamics, modifying the frequency of the oscillation while maintaining the edge modes isolated from the bulk.
Notice that in both cases the frequency of oscillations slightly shifts, which is expected due to the non-linear corrections produced by the interaction.

This difference in the exact dynamics for two particles confirms the method proposed in this work to engineer topological phases and provides a way to characterize them by detecting the time evolution of the charge occupation in the system.

Additionally, it is known that the edge states hosted by an SSH finite chain with non-trivial topology allow for long-range transfer of doublons directly from one end to the other without populating the intermediate region \cite{Miguel1}. Here we show that the two-electron states can be directly transferred between outer dots by considering topological models with a larger winding number. Then, the presence of more pairs of edge states, which can be controlled by choosing a suitable value for $\alpha/\omega$, opens up the possibility of designing new quantum-state-transfer protocols.\\

\textit{Conclusions}: We have proposed a driving protocol to engineer topological phases in a QD array with exponentially decaying hoppings. This is achieved by spatially modulating the driving amplitudes to imprint bond-ordering, and by selective enhancement or suppression of the different hopping processes. This generates the necessary symmetries for topological protection.
We have simulated a dimerized chain with chiral symmetry by setting stair-like driving amplitudes and dynamically quenching even hoppings. Furthermore, our protocol allows to enhance odd long-range hoppings versus short-range ones, thus opening the door to explore topological phases with different $W$. The use of square pulses allows for highly selective tunability of the different hoppings, where standard Floquet approaches using harmonic pulses would fail.

For the experimental implementation, scalable QD arrays of increasing size have been recently fabricated \cite{petta9qubits,vandersypen8}, making our proposal feasible with state of the art techniques.

To test our results we have simulated the exact dynamics for an initial product state, including local Coulomb interaction. We show that charge dynamics, which can be measured with quantum detectors in QD setups, discriminates between different topological phases. Additionally, we have found that the interplay of driving and interactions produces a drag effect between the electrons, which forms correlated edge modes; this is not only of fundamental interest but also relevant for quantum simulation and information purposes.

Importantly, our protocol can also be implemented in other platforms, or even straightforwardly extended to 2D systems. The main requirement is the local control of the driving amplitude at each site.
In optical lattices \cite{Gauthier16,Stuart2018} this could be done with additional lasers \cite{opticalAddressing}, and the engineering of long-range hoppings is well suited in this case by selection of certain optical transitions \cite{tunnelingOL}. In this setup, different topological features have been directly measured \cite{measurementZak,experimentoSSHcita, topOLexp,topOLexp2}.
Trapped ions can also be used, as it is possible to locally address each ion, and their effective Hamiltonian can be reduced to that of single excitations with long-range hopping decaying as $\sim d^{-3}$ \cite{diegoporrasDriving, porrasnevado}. Finally, molecular patterning on surfaces by adsorbates could also be considered \cite{STMfractalcris, STMKagomecris, STMliebcris, STMtop1, STMtop2}.\\

This work was supported by the Spanish Ministry of Economy and Competitiveness 
through Grant MAT2017-86717-P and we acknowledge support from CSIC Research Platform PTI-001. M. Bello acknowledges the FPI 
program BES-2015-071573, \'A. G\'omez-Le\'on acknowledges the Juan de la 
Cierva program and Beatriz P\'erez-Gonz\'alez acknowledges the FPU program FPU17/05297.
\bibliography{floquetbib}
\begin{widetext}

\section{SUPPLEMENTAL MATERIAL}

\subsection{Floquet Theory}

For driven periodic quantum systems with $H(t)=H(t+T)$, the presence of time translation symmetry enables the use of Floquet formalism. The Schrödinger equation can be solved in terms of the Floquet states $\ket{\psi_n(t)}=e^{-i\epsilon_n t}\ket{u_n(t)}$, where the so-called Floquet
modes $\ket{u_n(t)}=\ket{u_n(t+T)}$ have the same periodicity of the
Hamiltonian, and the role of static eigenenergies is assumed by the
quasienergies $\epsilon_n$.\\

In the high-frequency regime ($\omega\gg J_0$), the Floquet modes do not vary
much during a period compared with other frequency regimes. The dynamics is
essentially dictated by an effective time-independent Hamiltonian, which can be
derived as a series expansion in powers of $\omega$. In order to obtain an
effective Hamiltonian that is non-perturbative in the driving amplitudes $A_i$,
we first transform the original Hamiltonian in the rotating frame with respect to the
driving as 
\begin{equation}
\tilde{H}(t)=U^\dagger(t)H U(t)-iU^\dagger(t)\partial_tU(t)\,,
\end{equation}
with $U(t)=e^{-i \int dt\, H_\mathrm{driv}(t)}$, so that the new hopping
amplitudes change with a time-dependent complex phase 
\begin{equation}
\tilde{H}(t)=\sum_{i\neq j}J_{ij} e^{i(A_i-A_j)F(t)}c_i^{\dagger}c_j\,,
\end{equation}
and $F(t)=\int dt \, f(t)$. Now, the effective Hamiltonian can be obtained as
\begin{equation}
H_\mathrm{eff}=\sum_{\nu=0}^\infty \frac{1}{\omega^\nu}H_\mathrm{eff}^{(\nu)}
\,,
\label{eq:effHam}
\end{equation}
where the $H^{(\nu)}_\mathrm{eff}$ are functions of the components of the Fourier
decomposition of $\tilde{H}(t)$,
\begin{equation}
H_m=\frac{1}{T}\int_0^T \tilde{H}(t) e^{i\omega m t}dt\,.
\end{equation}
As it turns out, the leading term of this expansion is just the time-average of
the Hamiltonian, $H^{(0)}_{\mathrm{eff}}=H_0$ \cite{floquettheory, Mikami2016}. If the frequency 
is sufficiently large, truncating this expansion to zeroth
order gives already a good approximation to the effective Hamiltonian. The result is a time-independent Hamiltonian with renormalized hopping amplitudes.\\

\subsection{Gap}

The size of the gap has been calculated as a function of $\lambda$ and  $\alpha/\omega$. We aim to simulate configurations with a wide gap compared to the energy bands, so that the edge states are far from the extended states in the bulk to prevent them from hybridizing. As $|\alpha|/\omega$ is increased, the renormalized hoppings scale as $\omega/|\alpha|$, so the band structure is expected to shrink. This effect can be observed in Fig.\ref{fig:gap}, as the gap is bigger in the central region, corresponding to $\mathcal{W}=1$ and $\mathcal{W}=0$ phases and small values of $\alpha/\omega$.

\begin{figure}
	\includegraphics[width=0.5\linewidth]{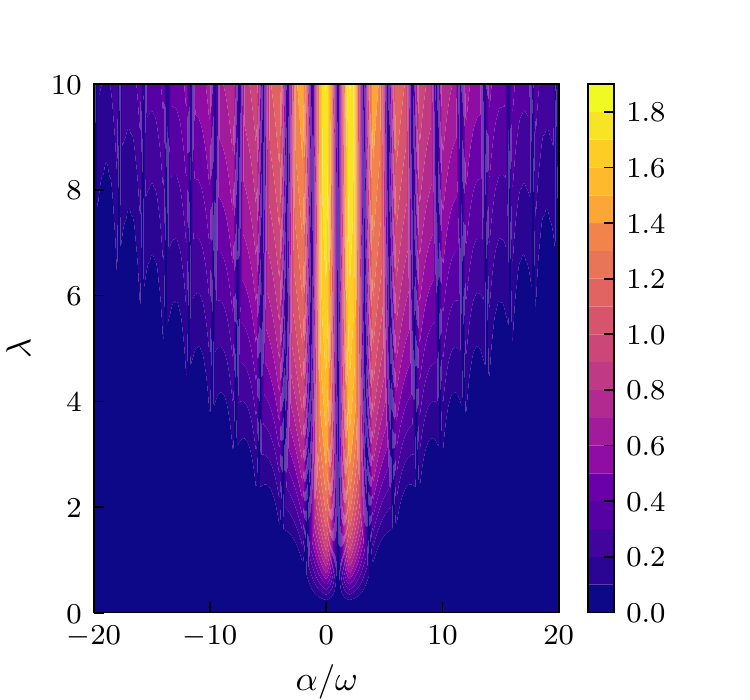}
	\caption{\label{fig:gap} Size of the gap as a function of the driving amplitude $\alpha/\omega$ and decay length of hopping amplitudes $\lambda$, for $q=1$. The maximum range of the hoppings included is fixed by $\lambda$, such that the longest hopping amplitude is a factor $10^{-8}$ smaller than the shortest one.}
\end{figure}

\subsection{Renormalized hopping amplitudes and Hamiltonian in the high-frequency regimen}

We consider a driven 1D quantum dot array in the high-frequency regime.
The system can be described through a time-independent Hamiltonian
with renormalized hopping parameters. Since the driving protocol achieves
the coherent destruction of all even-neighbor hoppings, we will only
consider odd-neighbor hoppings, which in turn are renormalized in
an alternating fashion. 

Hence, we can identify two types of renormalized hoppings. Being $r\in\mathbb{Z}_{\mathrm{odd}}$
the range of the hopping and $j$ the cell index, we first find that
hoppings of the form $J_{2j+r,2j}$ or $J_{2j,2j+r}$ (backward and
forward hopping, respectively) renormalize through

\begin{eqnarray}
A_{2j+r}-A_{2j} & = & \frac{r+1}{2}\beta+\frac{r-1}{2}\alpha=\left(r+1\right)\omega q-\alpha\\
A_{2j}-A_{2j+r} & = & -\left(A_{2j+r}-A_{2j}\right)=-\left[\left(r+1\right)\omega q-\alpha\right]
\end{eqnarray}

where we have made use of the relation $\alpha+\beta=2\omega q$ ($q=0,1,2,...$),
which is essential to imprint chiral symmetry. With this constraint,
$\beta$ is fixed by the value of $\alpha$ and thus we have only
one tuneable parameter. Therefore, these hopping amplitudes correspond
to

\begin{equation}
J_{2j+r,2j}\rightarrow\tilde{J}_{2j+r,2j}=J_{2j+r,2j}\frac{i\omega}{\pi\left[\left(r+1\right)q-\frac{\alpha}{\omega}\right]}\left(e^{-i\pi\left[\left(r+1\right)q-\frac{\alpha}{\omega}\right]}-1\right)
\end{equation}

\begin{equation}
J_{2j,2j+r}\rightarrow\tilde{J}_{2j,2j+r}=-J_{2j,2j+r}\frac{i\omega}{\pi\left[\left(r+1\right)q-\frac{\alpha}{\omega}\right]}\left(e^{+i\pi\left[\left(r+1\right)q-\frac{\alpha}{\omega}\right]}-1\right)
\end{equation}

On the other hand, hoppings of the form $J_{2j,2j-r}$ or $J_{2j-r,2j}$
(backward and forward hopping, respectively) renormalize through

\begin{equation}
J_{2j,2j-r}\rightarrow\tilde{J}_{2j,2j-r}^{\prime}=J_{2j,2j-r}\frac{i\omega}{\pi\left[\left(r-1\right)q+\frac{\alpha}{\omega}\right]}\left(e^{-i\pi\left[\left(r-1\right)\omega q+\frac{\alpha}{\omega}\right]}-1\right)
\end{equation}

\[
J_{2j-r,2j}\rightarrow\tilde{J}_{2j-r,2j}^{\prime}=-J_{2j-r,2j}\frac{i\omega}{\pi\left[\left(r+1\right)q-\frac{\alpha}{\omega}\right]}\left(e^{+i\pi\left[\left(r+1\right)\omega q-\frac{\alpha}{\omega}\right]}-1\right)
\]

In order to make an explicit distintion between both renormalized
hoppings, we have added a prime index on the second type. Since our
system has space-translation invariance, we can neglect the site index
in our notation and write

\begin{equation}
\tilde{J}_{2j+r,2j}\rightarrow\tilde{J}_{r},\hspace{1em}\tilde{J}_{2j,2j+r}\rightarrow\left(\tilde{J}_{r}\right)^{*}=\tilde{J}_{-r}
\end{equation}

\begin{equation}
\tilde{J}_{2j,2j-r}^{\prime}\rightarrow\tilde{J}_{-r}^{\prime},\hspace{1em}\tilde{J}_{2j-r,2j}^{\prime}\rightarrow\left(\tilde{J}_{-r}^{\prime}\right)^{*}=\tilde{J}_{r}^{\prime}
\end{equation}

Since hopping amplitudes are now complex functions, the direction
of the hopping is relevant and our notation must account for this
matter. We can see that both $\tilde{J}_{r}$ and $\tilde{J}_{-r}^{\prime}$
correspond to backward hoppings whereas their complex-conjugates $\tilde{J}_{-r}$
and $\tilde{J}_{r}^{\prime}$ correspond to forward hoppings.

In the high-frequency regimen, we can therefore describe the system
as bipartite and write the real-space Hamiltonian in terms of creation/annihilation
operators in each of the two sublattices $A$ and $B$. First, let
us keep the site indexes explicitely on the renormalized hopping parameters,
obtaining

\[
H=\sum_{r=1}^{R}\sum_{j=1}^{N}\left(\underbrace{\tilde{J}_{2j,2j-r}^{\prime}}_{\leftarrow}a_{j-\frac{r-1}{2}}^{\dagger}b_{j}+\underbrace{J_{2j+r,2j}}_{\leftarrow}b_{j}^{\dagger}a_{j+\frac{r+1}{2}}+\underbrace{\tilde{J}_{2j-r,2j}^{\prime}}_{\rightarrow}b_{j}^{\dagger}a_{j-\frac{r-1}{2}}+\underbrace{\tilde{J}_{2j,2j+r}}_{\rightarrow}a_{j+\frac{r+1}{2}}^{\dagger}b_{j}\right)
\]

where the arrows indicate the direction of the hopping.
The first (last) second terms correspond to backward (forward) hoppings.
Now, using the new notation for the hopping amplitudes,

\[
H=\sum_{r=1}^{R}\sum_{j=1}^{N}\left[\tilde{J}_{-r}^{\prime}a_{j-\frac{r-1}{2}}^{\dagger}b_{j}+\tilde{J}_{r}b_{j}^{\dagger}a_{j+\frac{r+1}{2}}+\left(\tilde{J}_{-r}^{\prime}\right)^{*}b_{j}^{\dagger}a_{j-\frac{r-1}{2}}+\left(\tilde{J}_{r}\right)^{*}a_{j+\frac{r+1}{2}}^{\dagger}b_{j}\right]
\]

The Hamiltonian in k-space can be written as

\begin{eqnarray}
H & = & \sum_{r=1}^{R}\sum_{k}\left(\tilde{J}_{-r}^{\prime}a_{k}^{\dagger}b_{k}e^{ikra}+\tilde{J}_{r}b_{k}^{\dagger}a_{k}e^{ikra}+\left(\tilde{J}_{-r}^{\prime}\right)^{*}b_{k}^{\dagger}a_{k}e^{-ikra}+\left(\tilde{J}_{r}\right)^{*}a_{k}^{\dagger}b_{k}e^{-ikra}\right)\\
& = & \sum_{r=1,k}^{R}\left(a_{k}^{\dagger},b_{k}^{\dagger}\right)\left(\begin{array}{cc}
0 & \tilde{J}_{-r}^{\prime}e^{ikra}+\left(\tilde{J}_{r}\right)^{*}e^{-ikra}\\
\left(\tilde{J}_{-r}^{\prime}\right)^{*}e^{-ikra}+\tilde{J}_{r}^{*}e^{ikra} & 0
\end{array}\right)\left(\begin{array}{c}
a_{k}\\
b_{k}
\end{array}\right)
\end{eqnarray}

where we have used $c_j=\frac{1}{\sqrt{N}}\sum_k c_k e^{ikx_j}$ and $c^\dagger_j=\frac{1}{\sqrt{N}}\sum_k c^\dagger_k e^{-ikx_j}$. The kernel of the Hamiltonian is

\[
\tilde{H}_{k}=\left(\begin{array}{cc}
0 & \tilde{J}_{-r}^{\prime}e^{ikra}+\left(\tilde{J}_{r}\right)^{*}e^{-ikra}\\
\left(\tilde{J}_{-r}^{\prime}\right)^{*}e^{-ikra}+\tilde{J}_{r}^{*}e^{ikra} & 0
\end{array}\right)
\]

and after performing an appropiated change of basis through $U=\mathrm{diag}\left(e^{ika},e^{-ika}\right)$,
we obtain

\[
H_{k}=U_{k}^{-1}\tilde{H}_{k}U=\sum_{r=1}^{R}\left(\begin{array}{cc}
0 & \tilde{J}_{-r}^{\prime}e^{ik\left(r-1\right)a}+\left(\tilde{J}_{r}\right)^{*}e^{-ik\left(r+1\right)a}\\
\left(\tilde{J}_{-r}^{\prime}\right)^{*}e^{-ik\left(r-1\right)a}+\tilde{J}_{r}^{*}e^{ik\left(r+1\right)a} & 0
\end{array}\right).
\]

We choose $a=\frac{1}{2}$ to resemble the Hamiltonian of the SSH
chain, finally resulting in

\[
H_{k}=\sum_{r=1}^{R}\left(\begin{array}{cc}
0 & \tilde{J}_{-r}^{\prime}e^{ik\frac{r-1}{2}}+\left(\tilde{J}_{r}\right)^{*}e^{-ik\frac{r+1}{2}}\\
\left(\tilde{J}_{-r}^{\prime}\right)^{*}e^{-ik\frac{r-1}{2}}+\tilde{J}_{r}e^{ik\frac{r+1}{2}} & 0
\end{array}\right).
\]

In terms of the Pauli matrices, $H_{k}$ can be written as $H_{k}=\vec{d}(k)\cdot\vec{\sigma}$,
with $\vec{d}(k)=\left(\mathrm{Re}\left[d\left(k\right)\right],-\mathrm{Im}\left[d\left(k\right)\right]\right)$
and $\vec{\sigma}=\left(\sigma_{x},\sigma_{y}\right)$, while $d\left(k\right)=$.
The topological invariant can be calculated as the winding number
of $\vec{d}\left(k\right)$ around the origin, defined as

\begin{equation}
\vec{d}\left(k\right)=\sum_{r=1}^{R}\left[\tilde{J}_{-r}^{\prime}e^{ik\frac{r-1}{2}}+\left(\tilde{J}_{r}\right)^{*}e^{-ik\frac{r+1}{2}}\right]
\end{equation}

\end{widetext} 
\end{document}